\documentstyle[epsf]{lamuphys}
\makeatletter
\let\chapter\hid@chapter
\makeatother
\begin{document}
\pagenumbering{arabic}
\title{Cluster Turbulence}

\author{Michael L.\,Norman\inst{1}\inst{,2} and Greg L.\,Bryan\inst{3}}

\institute{Astronomy Dept. and NCSA, University of Illinois, Urbana, 
IL 61801, USA
\and
Max-Planck-Institut f\"{u}r Astrophysik, D-85740 Garching, Germany
\and
Princeton University Observatory, Peyton Hall, Princeton, NJ 08544}
\maketitle

\def\etal{{\it et al.}}
\def\apj{ApJ}   
\def\mnras{MNRAS}
\def\aa{A\&A}
\def\newa{NewA}
\def\aj{AJ}

\begin{abstract}
We report on results of recent, high resolution
hydrodynamic simulations
\index{simulations, hydrodynamic}%
of the formation and evolution of X-ray clusters
\index{X-ray clusters}%
of galaxies carried out within a
cosmological framework. We employ the highly accurate piecewise
parabolic method (PPM)
\index{piecewise parabolic method}%
on fixed and adaptive meshes which
allow us to resolve the flow field in the intracluster gas. The
excellent shock capturing and low numerical viscosity of
PPM represent a substantial advance over previous studies using
SPH. We find that in flat, hierarchical cosmological models,
the ICM
\index{ICM}%
is in a turbulent state long after turbulence
\index{turbulence}%
\index{ICM!turbulence}%
generated by the last major merger should have decayed away. 
Turbulent velocites are found to vary slowly with cluster radius,
being $\sim 25\%$ of $\sigma_{vir}$ in the core, increasing to 
$\sim 60\%$ at the virial radius. We argue that more frequent
minor mergers maintain the high level of turbulence found in the
core where dynamical times are short. Turbulent pressure support is 
thus significant throughout the cluster, and results in a somewhat
cooler cluster ($T/T_{vir} \sim .8$) for its mass. 
Some implications of cluster turbulence are discussed.
  
\end{abstract}
\section{Introduction}
Our conception of galaxy clusters\footnote{to appear in 
{\em Ringberg Workshop on M87}, eds. K. Meisenheimer
\& H.-J. R\"{o}ser, Springer Lecture Notes in Physics, 1998.}
as being dynamically 
relaxed systems has undergone substantial revision in recent years.
Optical observations reveal substructure
\index{clusters!substructure}%
in 30-40\% of rich
clusters (Geller \& Beers 1982; Dressler \& Shectman 1988). 
A wealth of new X-ray observations have bolstered these findings,
providing evidence of recent mergers in clusters previously
thought to be archtypal relaxed clusters (e.g., Briel \etal ~1991).
Also eroding the conventional view has been the success of
``bottom-up" or hierarchical models
\index{hierarchical models}%
of cosmological structure formation in accounting for the formation of
galaxies and large scale structure
\index{galaxies!formation}%
\index{galaxies!large scale structure}%
\index{galaxies!X-ray clusters}%
in the universe (e.g., Ostriker 1993).
Within such models, a cluster sized object is built up through
a sequence of mergers of lower-mass systems (galaxies $\rightarrow$ 
groups $\rightarrow$ clusters). 
In a flat universe ($\Omega_o = 1$) as
predicted by inflation,
\index{inflation}%
mergers would be ongoing at the present 
epoch. In open models ($\Omega_o < 1$), 
mergers cease at a redshift $z \sim \Omega_o^{-1}-1$, and 
clusters become relaxed by today. The amount of substructure observed
in X-ray clusters of galaxies at $z \sim 0$ is thus a powerful probe of
cosmology. Evrard \etal ~(1993) and Mohr \etal ~(1995)
have explored this ``morphology-cosmology" connection,
\index{morphology-cosmology connection}%
and concluded that a high $\Omega$ universe is favored. 
Interestingly, Tsai \& Buote (1996) reach the opposite conclusion.

Cluster mergers
\index{cluster mergers}%
\index{simulations!cluster mergers}%
have been explored numerically by several 
groups (Schindler \& M\"{u}ller 1993; Roettiger, Loken \& Burns 1997;
Roettiger, Stone \& Mushotzky 1998). In these hydro/N-body simulations,
two hydrostatic King models are collided varying the cluster--subcluster
mass ratio. It is found that major mergers 
induce temperature inhomogeneities and
bulk motions in the ICM of a substantial fraction 
\index{ICM, bulk motions}%
of the virial velocity ($> 1000 ~km/s$). Roettiger
\etal ~suggest that these bulk motions may be responsible for
the observed temperature substructure seen in some X-ray clusters,
as well as bending Wide-Angle Tailed radio sources,
\index{WAT radio sources}%
energizing cluster radio halos,
\index{clusters!radio halos}%
and disrupting cooling flows.
\index{clusters!cooling flows}%

If hierarchical models are correct, the thermal and dynamical
state of the ICM could be considerably more complex than the above
mentioned simulations indicate. 
In a flat universe, for example, the ICM would be constantly bombarded by
a rain of minor mergers in addition to the occasional major merger. 
Also omitted in those simulations are
a variety of cosmological effects which may be important, including
memory of the complex formation history of the merging clusters,
infall of matter along filaments, accretion shocks, 
large scale tides, and cosmic expansion. 

In this paper we show results of numerical simulations that take
all these effects into account. We find in two flat models investigated,
that quite generally the ICM of rich galaxy clusters
is in a turbulent state. The turbulent velocities are typically 
60\% the virial velocity at the virial radius, decreasing inward
to roughly 25\% within the core. The relatively slow decline
in turbulence amplitude with decreasing radius suggests that
frequent minor mergers are an important driving mechanism
in addition to rare massive mergers.  
In addition, we find ordered fluid circulation in the core of
one well--resolved cluster which is likely the remnant of
a slightly off-axis recent merger.

\section{Simulations}

The simulations are fully cosmological. That is, the formation and
evolution of the clusters is simulated by evolving the equations of
collisionless dark matter,
\index{dark matter}%
primordial gas and self-gravity in an
expanding FRW universe (see e.g., Anninos \& Norman 1996).
\index{FRW universe}%
Initial conditions consist of specifying
linear density and velocity perturbations in the gas and dark matter
in Fourier space 
with power spectrum $P(k)$ and random phases. We have simulated two
cosmological models which differ primarily in their assumed $P(k)$'s: 
CDM, with power normalized to reproduce the abundance of great clusters 
\index{CDM}%
at z=0, and CHDM, 
\index{CHDM}%
normalized to the COBE
\index{COBE}%
measurement on large scales.
We assume the gas is non-radiative, which is a good approximation 
except in the cores of cooling flow clusters. 
The statistical properties
of X-ray clusters in these models (as well as an open model) are 
presented in Bryan \& Norman (1998a). The internal structure of a smaller
sample of X-ray clusters computed at higher resolution are presented
in Bryan \& Norman (1998b). Here we summarize the key findings from
Bryan \& Norman (1998b), restricting ourselves to the properties of
four clusters drawn from the CDM simulations. Table 1 summarizes the 
clusters' bulk properties. 

Two different numerical gridding techniques have been employed. 
The first uses a uniform Eulerian grid with $512^3$ cells in a comoving
volume of 50 Mpc, for a cell resolution of $\sim 100 kpc$. While unable 
to resolve the cluster core, this calculation provides good       
coverage in the cluster halo and beyond. Three clusters, called CDM1-3,
are taken from this simulation. The second employs adaptive
mesh refinement (AMR; Bryan \& Norman 1997a) 
\index{adaptive mesh refinement}%
which automatically adds high resolution
subgrids wherever needed to resolve compact structures, such as 
subclusters forming at high redshift or the cluster core at $z=0$.
We have computed a single rich cluster, called SB, with 15 kpc resolution
in the core (Bryan \& Norman 1997b).
This cluster has been simulated by a dozen groups in the ``Santa Barbara
cluster comparison project", (Frenk \etal ~1998). 
Together, these simulations allow us to characterize the
thermal and dynamical state of the ICM across a wide range of scales.
Both simulations use the piecewise parabolic method (PPM) for gas dynamics,
\index{piecewise parabolic method}%
modified for cosmology (Bryan \etal ~1995), and the particle-mesh method
(PM) for the dark matter dynamics. 
The excellent shock capturing and low numerical viscosity of
PPM make it ideal to study cluster turbulence. 

% +++++++++++++ Table 1 +++++++++++++

\begin{table}
\centering
\begin{tabular}{|c|c|c|c|c|c|}
\hline
cluster & $r_{vir}$ (Mpc) & $M_{vir}$ ($10^{15} M_\odot$) &
$T_{vir}$ (keV) & $\sigma_{vir}$ (km/s) & $\Delta x$ \\
\hline
\hline
CDM1 & 2.58 & 0.890 & 4.63 & 861 & 98 \\
CDM2 & 2.32 & 0.647 & 3.74 & 774 & 98 \\
CDM3 & 2.40 & 0.716 & 4.00 & 801 & 98 \\
SB & 2.70 & 1.1   & 4.71 & 915 & 15 \\
\hline
\end{tabular}
\newline
\caption{Cluster parameters}
\label{table:virial_quantities}
\end{table}

\index{virial velocity}%
\index{virial radius}%
\index{virial temperature}%
% +++++++++++++++++++++++++++++++++++

\section{Turbulence in the Halo}
\index{clusters!halo}

\begin{figure}
\epsfysize=4in
\centerline{\epsfbox{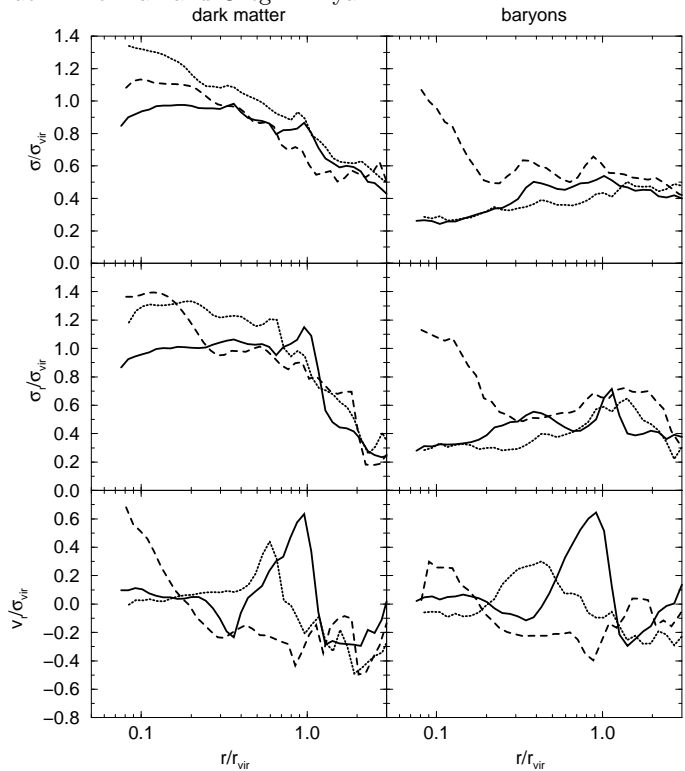}}
\caption{
The velocity dispersion (top panels), radial velocity dispersion
(middle panels) and radial velocity in shells for the dark matter
(left side) and gas (right side) of the three largest clusters in the
CDM512 simulation (solid/dotted/dashed lines correspond to clusters
designed as CDM1/CDM2/CDM3).  Profiles are normalized by their virial
values (see Table 1).
}
\label{fig:sigma_6panel_profiles}
\end{figure}

In Figure~\ref{fig:sigma_6panel_profiles}, we plot the azimuthally
averaged total velocity
dispersion, radial component of the velocity dispersion and the radial
velocity for both the dark matter and gas components of 
clusters CDM1-3.  They are normalized by the virial values from
Table~\ref{table:virial_quantities} and all velocities are relative
to the center-of-mass velocity of the matter within $r_{vir}$.  The
dispersion in the radial direction is around the net radial velocity
of that shell: $\sigma_r^2 = \langle (v_r - \langle v_r \rangle)^2
\rangle$. 

% ++++++++++++ sigma_6panel profile +++++++++++

% +++++++++++++++++++++++++++++++++++
% +++++++++++++++++++++++++++++++++++

\begin{figure}
\epsfysize=4.5in
\centerline{\epsfbox{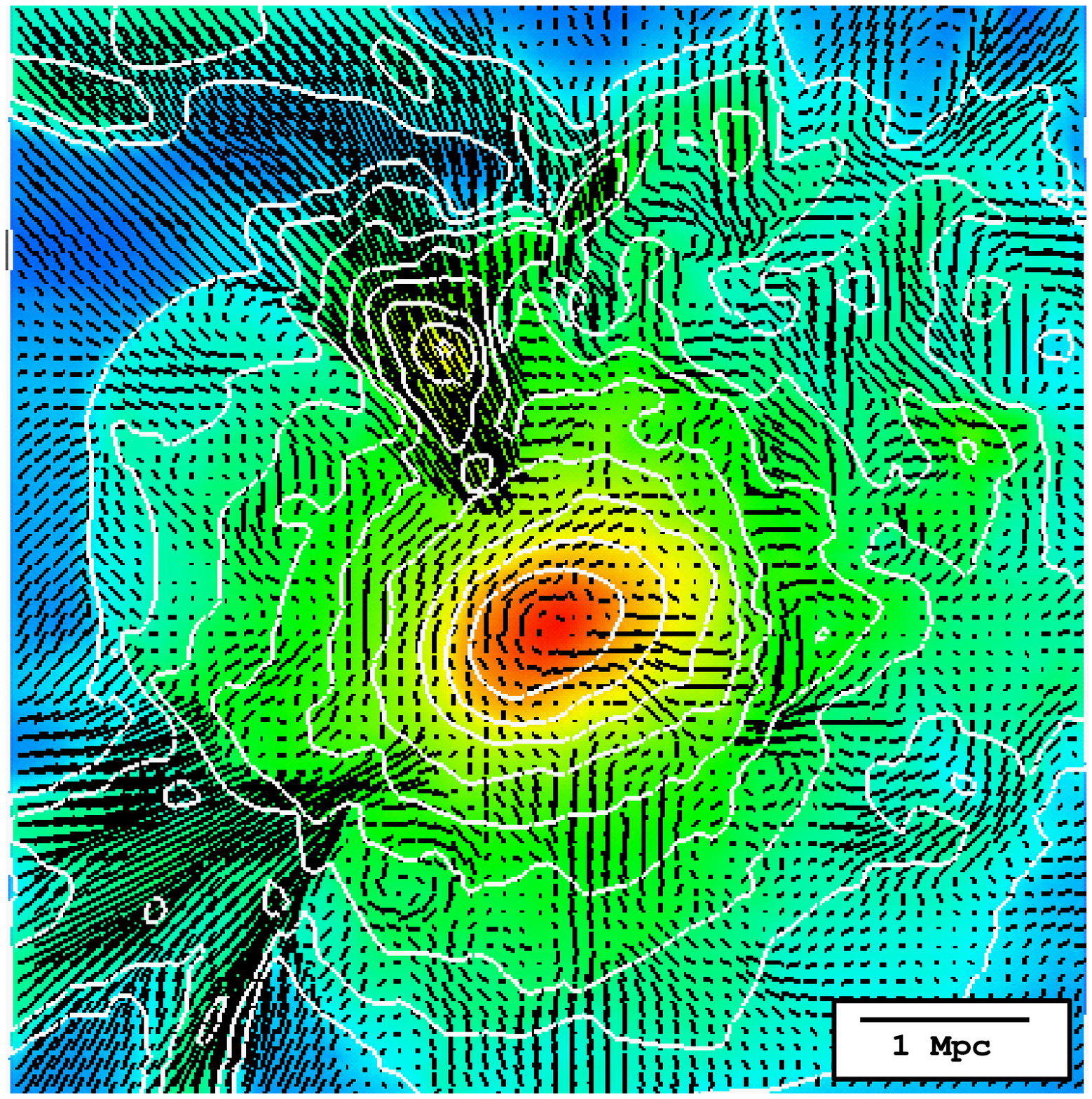}}
\caption{
The large scale velocity field on a thin slice though the center of 
cluster SB shown overtop the logarithm of gas density (image, contours).
The maximum velocity vector is 2090 km/s. The image is 6.4 Mpc on a side.
}
\label{fig:XY_slice_6.4Mpc}
\end{figure}

% +++++++++++++++++++++++++++++++++++

Focusing first on the dark matter, the velocity dispersion profiles
are roughly compatible with their virial values within the virial
radius, but fall off quickly beyond that point.  There is some
preference for radial orbits around and slightly beyond $r_{vir}$, but
at low radii, the velocities are isotropic.  The radial velocity
profile (bottom panel) shows evidence for infall in the 1--4$r_{vir}$
range. The third cluster in this sample (dashed line) is undergoing 
a major merger and
shows signs of enhanced bulk motions in the inner 400 kpc, although
the velocity-dispersion profiles are not strongly disturbed.
\index{velocity dispersion}%

The gas velocity dispersions range between 0.25 and 0.6 $\sigma_{vir}$,
considerably below their dark matter counterparts, but are not
insignificant.  In fact, these motions contribute some additional
support beyond that provided by the mean baryonic pressure gradient.
We may approximate this by appling Jean's equation
\index{Jean's equation}%
to the coherent
clumps of gas with velocity dispersion $\sigma$ and density $\rho_c$.
Ignoring differences between the radial and tangential velocity
dispersion this becomes:
\begin{equation}
\frac{1}{\rho_{c}} \frac{d (\rho_{c} \sigma^2)}{dr} +
%	2\frac{\sigma_r^2 - \sigma_t^2}{r}      + 
	\frac{1}{\rho}\frac{d P}{d r}
	= - \frac{GM(r)}{r^2}.
\label{eq:jeans_like_equation}
\end{equation}
Since $P = \rho kT/\mu m_h$, where $\mu m_h$ is mean mass per
particle, we see that the temperature and $\sigma^2$ combine to
support the cluster gas against gravitational collapse.  We can
directly compare $T/T_{vir}$ against $\sigma^2/\sigma^2_{vir}$, so the
temperature provides about 80\% of the support.  This provides an
explanation for the observation (Navarro, Frenk \& White 1995; Bryan
\& Norman 1998a) that the mean cluster temperatures were, on average, 
about $0.8$ of its virial value.

Thus we see that the gas has not completely virialized and sizable
bulk motions exist.  Since the mean entropy profile increases with
increasing radius, the halo is globally stable, so this turbulence
must be driven by external masses falling into the cluster and damped
by viscous heating.  The turbulence amplitude in the halo
appears to be roughly
compatible with this explanation since the driving timescale ---
approximately the Hubble time --- is slightly larger than the damping
timescale which is essentially the crossing time.  Moreover,
$\sigma^2$ seems to drop (and $T$ approaches $T_{vir}$) as $r
\rightarrow 0$ and the crossing time decreases. We discuss this point 
further in the last section.

Figure~ \ref{fig:XY_slice_6.4Mpc} shows the chaotic flowfield on a
slice through the center of cluster SB.
Velocity vectors for the gas are superposed on the log of the gas density.
High velocity streams seen at 8 and 11 o'clock are caused by inflow of
low entropy material along large scale filaments. This low entropy gas 
sinks to the center of the cluster. Generally, subclusters fall in
along filaments, and their passage through the cluster generates
vorticity, seen here as large scale eddies, 
via the baroclinic mechanism
\index{vorticity, generation}%
(e.g., Stone \& Norman 1992).
The eddies are $\sim 500 ~kpc$
in diameter and have a velocity of $\sim 1000 ~km/s$.
Between the filaments, gas can actually move outwards. In this cluster,
a portion of the main accretion shock
\index{shock waves}%
is visible in the upper left corner.
The infalling gas impacts the shock with a
range of angles.  When the velocity is normal to the shock front, the
gas is almost completely virialized, however, oblique impacts generate
substantial vorticity in the post-shock gas.  This is another source
of turbulent motions in the cluster gas.

\section{Bulk Motions in the Core}
\index{clusters!core}%

% +++++++++++++++++++++++++++++++++++

\begin{figure}
\epsfysize=4.0in
\centerline{\epsfbox{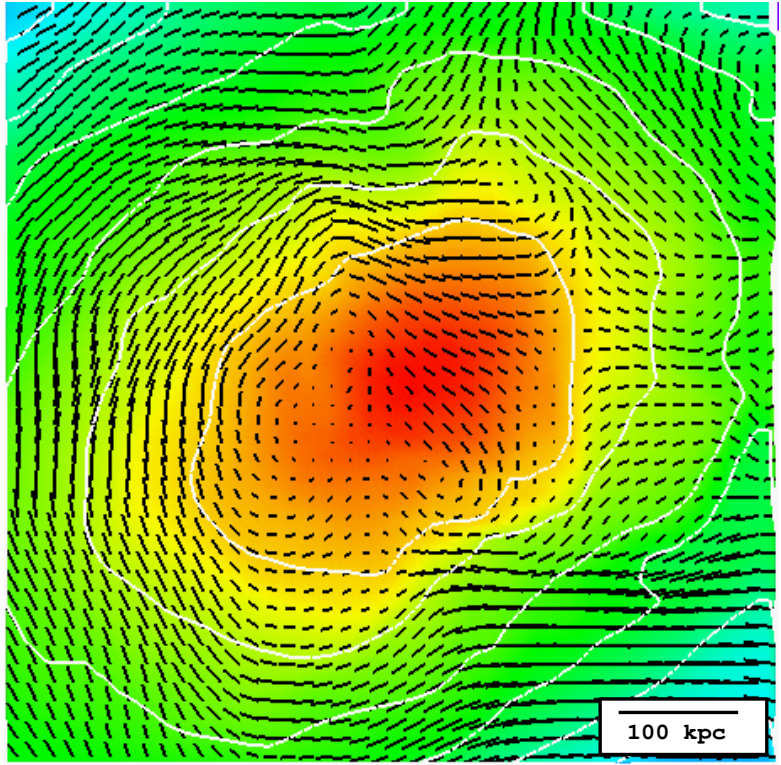}}
\caption{
The velocity field on a thin slice 
in the inner 600 kpc of cluster SB. The maximum velocity
vector is $520 ~km/s$.
}
\label{fig:core_vel_slice}
\end{figure}

% +++++++++++++++++++++++++++++++++++

Inside $1 ~Mpc$, we see coherent bulk motions with typical velocities of
$\sim 500 ~km/s$ and correlation scales between $100 ~kpc$ and $1 ~Mpc$.
The geometry of the flow is complex, changing character
on different slices.
The slice shown in Fig. ~\ref{fig:XY_slice_6.4Mpc}
shows a large-scale circulation about the cluster core.

Using the high resolution model SB, we can probe the
velocity field on scales down to $0.01 r_{vir} = 27 ~kpc$. 
A blow up of the central portion of the cluster
shown in Fig. ~\ref{fig:core_vel_slice}. Here, the spacing of the vectors
corresponds to our cell size $15 ~kpc$. The clockwise circulation is clearly
evident here, and numerically well resolved. Within the central $200 ~kpc$,
we can see eddies 4-5 cells in diameter---close to
our resolution limit. 
Thus, turbulence exists even in the cores of X-ray clusters.

% +++++++++++++++++++++++++++++++++++
\begin{figure}
\epsfysize=4.0in

\centerline{\epsfbox{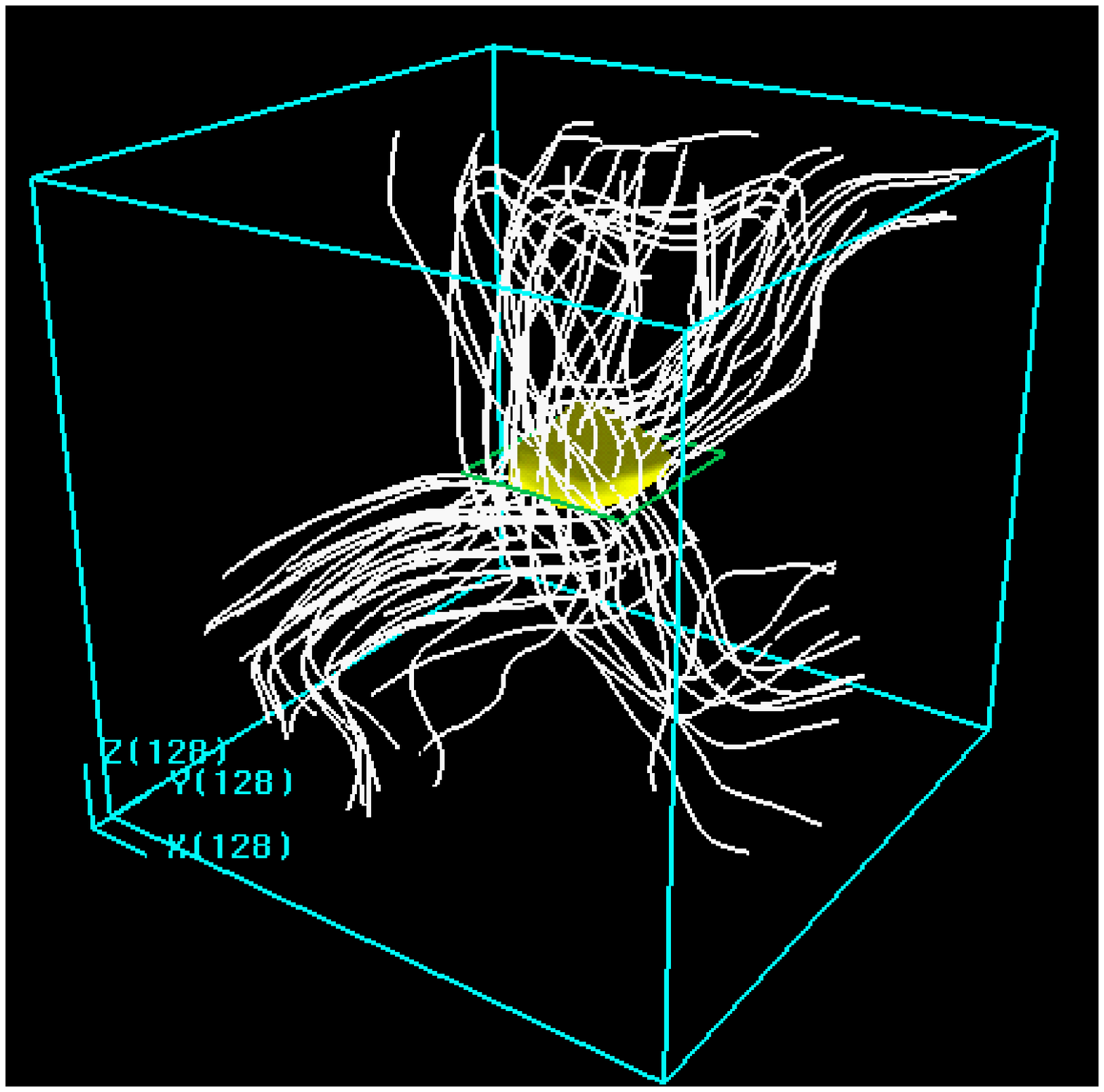}}
\caption{
The 3d velocity field in the inner 600kpc centered on the core.
}
\label{fig:streaklines}
\end{figure}
% +++++++++++++++++++++++++++++++++++

Fig. ~\ref{fig:streaklines} shows the three dimensional velocity field
in a $600 ~kpc$ box centered on the core (shaded isosurface). 
We find that the flow is quite ordered on these scales, with bulk
velocities of $300-400 ~km/s$. Here we have rendered fluid
``streaklines" passing through the core, which are tangent curves of
the instantaneous velocity field. In a steady flow, streaklines and
streamlines are identical and trace out the paths that fluid elements
follow. In a time-dependent flow, such as we have
here, streaklines provide only a sense of the geometry of the velocity 
field. Close inspection of Fig. ~\ref{fig:streaklines} as well as 3d rotations
on graphics workstation reveal a swirling flow superposed on a linear
flow. The linear flow corresponds to the mean peculiar velocity of the 
cluster core, which points from the origin of the cube to the upper right
furthest corner of the cube. The swirling flow can be seen as the 
bundle of streaklines coming out of the page below the core, passing
in front of the core, and going back into the page above the core.

\section{Discussion}

We have shown using high resolution hydrodynamic simulations
that the ICMs in bright X-ray clusters in flat hierarchical
models are turbulent throughout.
The turbulence in strongest in the outskirts of the cluster
and weaker in the core. 
Due to the declining temperature profile in cluster halos,
the turbulence is found to be mildly supersonic
($M \sim 1.6$) near $r_{vir}$, decreases rapidly to $M \sim 0.5$
at $\sim \frac{1}{3}r_{vir}$, and thereafter declines more slowly 
to $M \sim .3$ in the core. 

Here we argue that infrequent major mergers cannot sustain the
observed level of turbulence in the core. It is known from
simulations of decaying turbulence in a box that the turbulent
kinetic energy decays as $t^{-\eta}$ where t is measured in 
units of the dynamical time.
\index{turbulence!decay rates}%
The exponent $\eta$ depends weakly on the nature of the
turbulence, but is around 1.2 for compressible,
adiabatic, hydrodynamic turbulence (Mac Low \etal ~1998). 
The time for a sound wave to propagate from the center of the
cluster SB to a radius $.01, .1, 1 \times r_{vir}$ is
$.014, .173, 3.1 ~Gyr$, respectively. The cluster
underwent a major merger at $z=0.4$, or $5.2 ~Gyr$ earlier. 
Taking the sound crossing time as the dynamical time, we
predict that fluid turbulence induced by the major merger 
at $z=0.4$ would have decayed to $.006, .017, .56$ of
its initial value by $z=0$.

Several possibilities suggest themselves to account for the 
high fluid velocity dispersions seen in the core. The first
is that energy is somehow pumped 
into the core by motions in the outer parts of the cluster
which relax on longer timescales. However, shock waves
generated by supersonic motions in the outskirts would
weaken into acoustic disturbances as they propagated into
the dense, hotter core. Gravitational accelerations in the
core would be dominated by the local dark matter distribution
which would relax on a timescale comparable to the turbulence
decay timescale. Another pumping mechanism discussed by 
Roettiger, Burns \& Loken (1996) 
is global oscillations of the cluster potential following 
a major merger. They find that rms velocities decay to 
$\sim 200 ~km/s$ by 2 Gyr after core passage, and remain 
quite constant thereafter. This is substantially less than 
the velocities we find.  

The second possibility, which we consider
more likely, is that core turbulence is driven by the more
frequent minor mergers. 
Lacey \& Cole (1993) have quantified the merger rates 
in hierarchical models. They find that the merger rate
for CDM scales as $(\Delta M/M_{cl})^{-\frac{1}{2}}$
where $\Delta M$ is the subcluster mass. Whereas most
of a cluster's final mass is typically accreted in a single major
merger, they find that the cluster will typically accrete $\sim 10\%$ 
of its mass in ten minor mergers of clumps $\sim 1\% $ of its final mass.
The most probable formation epoch for a $10^{15} M_{\odot}$ 
cluster in the stardard CDM model we have simulated 
is at .7 $t_{Hubble}$, or 4 Gyr ago. 
The mean time between minor mergers is thus 0.4 Gyr---comparable
to the dynamical time at a tenth the virial radius.

Is there sufficient energy in minor mergers to sustain the
turbulence in the core, and if so, how is the energy deposited?
The kinetic energy of ten $10^{13} M_{\odot}$
subclusters is $\sim 10^{63}$ erg, as compared to approximately
$10^{62}$ erg of turbulent kinetic energy within $0.1 r_{vir}$.
Thus, a 10 \% energy conversion efficiency is required
for this mechanism to be correct. 
If the coupling is purely hydrodynamic (i.e., shocks), then
the energy available is the kinetic energy of the gas in the
subcluster, which is down by a factor of 
$\Omega_b$ from the estimate above.
Since $\Omega_b =.05 $, this energy is insufficient. Thus,
it would seem that a substantial gravitational coupling
between the ICM and the dark matter in the subclusters is
required. This is equivalent to saying that the gas remains
bound to the subcluster until it reaches the core. 
Roettiger \etal ~(1996) found that this is indeed the case.

There are a number of interesting implications to significant levels 
of turbulence in the cores of X-ray clusters, many of which
have already been pointed out by Roettiger \etal ~(1996),
including Doppler shifting of X-ray emission lines, bending
of Wide-Angle Tailed radio galaxies, and powering cluster radio
halos. Our findings strengthen their conclusions.
For example, the turbulent amplification of magnetic fields
would be expected to be most efficient in cluster cores where 
dynamical timescales are shortest.
Moreover, continuous stirring by minor mergers 
could modify cooling flows appreciably. Because 
turbulent pressure ``cools" inefficiently compared to atomic
processes, turbulent pressure 
support could become increasingly important in the central parts
of a cooling flow. 
Its effect would be to reduce the mass inflow rate into
the cluster center. Secondly, at radii much less than the
cooling radius, turbulent motions would concentrate cooling
gas into filaments, and possibly account for the observed
$H \alpha$ filaments. Finally, we note that ordered
circulation in the cores of X-ray clusters such as we have found
might account for the S-shaped symmetry of radio tails seen
in some sources (e.g., M87; B\"{o}hringer \etal ~(1995), 
Owen, these proceedings.)
\index{M87!radio tails}%

{\it Acknowledgements:} This work was partially supported by
grants NASA NAGW-3152 and NSF ASC-9318185. Simulations were
carried out on the Connection Machine-5 and Silicon Graphics
Power Challenge Array at the National Center for Supercomputing
Applications, University of Illinois.
%

%
% ---- Bibliography ----
%

\end{document}